\documentclass[aip,reprint]{revtex4-1}

\draft 

\begin{document}


\title{On the conjectured gravity-related collapse rate $E_\Delta/\hbar$
       of massive quantum superpositions} 



\author{Lajos Di\'osi}
\email[e-mail: ]{diosi.lajos@wigner.hu}
\homepage[homepage: ]{www.wigner.hu/~diosi}
\affiliation{Wigner Research Center for Physics, Budapest, Hungary}
\altaffiliation{Also at Department of Physics of Complex Systems, 
E\"otv\"os Lor\'and University, Budapest, Hungary}


\date{\today}

\begin{abstract}
Roger Penrose and the author share the proposal that the spatial superposition 
$|x_1\rangle+|x_2\rangle$ of a massive object collapses into its localized components 
$|x_1\rangle$ or $|x_2\rangle$ with the characteristic time $\hbar/E_\Delta$ where $E_\Delta$ 
is the gravitational self-energy excess of the superposition versus the localized states.  
Underlying arguments of such radical  departure from standard quantum mechanics and 
different derivations of the rate equation are briefly recapitulated and discussed. 
\end{abstract}


\maketitle 


\section{The collapse rate}
\label{Rate}Microobjects, from elementary particles to giant molecules, can exist in 
superpositions of different locations. As to more massive objects, however, the 
violation of standard quantum mechanics has been conjectured from certain purely 
theoretical speculations.
The proposal concerned here has been surviving three and a half decades in a status
of pure speculation. The overlap between Penrose's and my results is the claim
that  the spatial superposition $|x_1\rangle+|x_2\rangle$ becomes unstable 
for large masses and a random collapse
\begin{equation}
\label{collapse}
|x_1\rangle+|x_2\rangle
\Rightarrow\left\{\begin{array}{ll}|x_1\rangle&\mbox{ with probability }0.5\\
                                   |x_2\rangle&\mbox{ with probability }0.5
                                    \end{array}\right.
\end{equation}
happens at rate
\begin{equation}
\label{rate}
\frac{1}{\tau}=\frac{E_\Delta}{\hbar}~,
\end{equation}
where $\tau$ is the mean lifetime of the superposition and $E_\Delta$ is
the difference between gravitational self-energies before and after
the collapse (\ref{collapse}), respectively, times an unspecified numeric constant.
 
After decades of missing experimental evidences pro or contra, 
the advent of quantum controlled laboratory technique
opened the era of testability. It is worthwhile to revisit the 
theoretical background, the  diverse arguments that seem 
to converge to the above collapse rate.  
 
Secs. \ref{Penrose},\ref{Diosi} attempt to outline the proposals of Penrose
and myself, respectively. Sec.\ref{Discussion} discusses my occasional selection
of related issues, followed by closing remarks in Sec.\ref{Closing}. 
 
\section{Collapse from conjectured Killing vector ambiguity}
\label{Penrose}
Let me try biefly interpreting Penrose's concept  and arguments
\cite{penrose1994shadows,penrose1996,penrose2002},
leading him to the rate (\ref{rate}). 
 
Consider the center of mass stationary state $|x_1\rangle$ of a massive object 
located at $x_1$ and the stationary state $|\gamma_1\rangle$ of the geometry 
corresponding to the state $|x_1\rangle$. 
The composite state $|x_1\rangle\otimes|\gamma_1\rangle$ 
is also stationary. Now take the same stationary state just shifted from $x_1$ 
to $x_2$ and consider the superposition:
\begin{equation}
\label{superposition}
|x_1\rangle\otimes|\gamma_1\rangle+|x_2\rangle\otimes|\gamma_2\rangle~.
\end{equation}
One would expect that this superposition is also stationary. 
Penrose argues that it can not be. Independently of the details of how $\gamma_1$
and $\gamma_2$ represent the two geometries, they have their own Killing
vectors to define stationarity but they have no single common Killing vector to
define stationarity of the superposition.  The equivalence principle of general relativity
(general covariance, in other terms) 
\emph{``forbids a meaningful precise labelling of individual points in a space-time.
[...] there is generally no precise meaningful pointwise
identification between different space-times''} --- says Penrose\cite{penrose1996} and 
adds: \emph{``all that we can expect will be some kind of approximate pointwise
 identification''}.

The \emph{``measure of this degree of approximation''} is obtained by Penrose in 
the Newtonian non-relativistic limit of general relativity. 
The time coordinate for the two geometries $\gamma_1,\gamma_2$ can 
now be taken the common $t$, the Killing vector becomes equivalent to
``$\partial/\partial t$'' while it remains ambiguous because the point-wise
identification of the spatial coordinates $x$ remains ambiguous.
Penrose argues that this ambiguity corresponds to the ambiguity of free falls
determined by the ambiguity of local accelerations $g=-\nabla\Phi$ where
$\Phi$ is Newton's potential.
If so, then the plausible measure of the ambiguity (uncertainty) will be
proportional to the volume integral of the squared difference of local accelerations:
\begin{equation}
\label{uncertaintyPhi}
\Delta \propto \int \left\vert g_1(x)-g_2(x)\right\vert^2 d^3x~.
\end{equation}
One expresses $g_1$ by the mass distribution $\rho_1$  in 
state $|x_1\rangle$:
\begin{equation}
g_1=-\nabla\Phi(x)=-G\nabla\int\frac{\rho_1(x')}{|x-x'|}d^3x'~,
\end{equation}
and similarly for $g_2$. The ambiguity, or ``uncertainty'' $\Delta$, divided by $G$,
takes this form:
\begin{eqnarray}
\label{EG}
E_\Delta\!&=&\!\mbox{const.}\!\times \!G\!\!\int\!\!\!\!\int\!
\frac{\left(\rho_1(x)\!-\!\rho_2(x)\right)\!\left(\rho_1(x'\!)-\!\rho_2(x')\right)}
     {|x-x'|}d^3xd^3x'\nonumber\\
         &=&\!\mbox{const.}\!\times\left(U(x_1-x_2)-U(0)\right)~,
\end{eqnarray}
where $U(x_1-x_2)$ is the Newton interaction potential between 
$\rho_1$ and $\rho_2$.

The bottom line of the derivation is that the energy $E_\Delta$ should be 
considered the energy ambiguity of the superposition (\ref{superposition}) and,
as for usual unstable quantum states, $E_\Delta$ leads to decay at mean lifetime 
$\tau$ defined in (\ref{rate}). 

Later, Penrose supports the role of acceleration $g$ in the uncertainty measure
(\ref{uncertaintyPhi}) by an alternative reasoning \cite{penrose2014}. 
Consider a mass $M$ in free fall and compare its wavefunctions in the
Earth system and in the free-falling system, respectively. The former has a
phase factor $\exp(-iMgt^3/6\hbar)$.  A key part of the new arguments 
invokes relativistic quantum field theory where the two wavefunctions
would belong to two different vacua, i.e., to non-equivalent Hilbert spaces.
Although alternative vacua are irrelevant in the given non-relativistic situation,
as noticed by Penrose, his train of thought may still hit the target, despite
the overlooked triviality of the phase factor. The strange-looking phase 
comes simply from the nonzero time-dependent kinetic energy in the Earth 
based frame:  $Mgt^3/6$ is the integral of $Mgt^2/2$.

\section{Collapse from conjectured geometric ambiguity}
\label{Diosi}
My approach goes like this.
Curvature of space-time geometry is sourced by the energy-momentum 
of matter which is quantized obviously. Hence quantum uncertainties of matter's
behavior should impose uncertainties of geometry as well. 
This unsharpness of geometry is thus unavoidable and depends on 
$\hbar$, but its details depend on the model that couples quantized matter
and quantized (or perhaps classical) geometry. Independently of the model,
we might nonetheless estimate the scale of uncertainties transferred from 
matter to gravity. The concept is this. The uncertainty of the geometry coincides 
with the optimum testability of geometry, using quantized material instruments. 
In particular, considering a network of quantized free falling test bodies to 
measure the geometry, one expects that there is a finite optimum of measurement 
precision. 

The measure of this precision and the rate (\ref{rate}) is obtained in 
the Newtonian non-relativistic limit of general relativity 
\cite{diosi1986thesis,diosilukacs1987}.
Let us analyse how precisely the free fall  of a single test mass $M$ encodes the
the local acceleration $g=-\nabla\Phi$.
Let the standing initial wave packet of $M$ have a certain size $\sim r$ and volume
$V\sim r^3$. Under free fall, $r$ is approximately retained over a period 
$T\sim Mr^2/\hbar$. Hence, the test mass encodes the average acceleration field
$\bar g $ over the volume $V$ and time $T$.  The acquired momentum   
$M\bar g T$, part of the total one,  has an uncertainty $\hbar/r$. Hence $\bar g $ is
encoded at the precision
\begin{equation}
\delta\bar g \sim\frac{\hbar}{MrT}~.
\end{equation}
To improve precision, one can not increase $M$ boundlessly because $M$'s Newton 
potential contributes to $\bar g $ and imposes a further uncertainty
\begin{equation}
\delta\bar g \sim\frac{GM}{r^2}~,
\end{equation}
because of $M$'s position uncertainty $r$. The optimum value of the test
mass $M$ is reached when the above two uncertainties coincide. 
Then the  optimum precision of the measurement reads:
\begin{equation}
\label{uncgVT}
\delta\bar g \sim\sqrt{\frac{\hbar G}{VT}}~.
\end{equation}
The factor $1/\sqrt{VT}$ suggests that the uncertainties of $g$ at different locations 
and different times are independent. 
One can determine the corresponding structure and scale of 
uncertainties $\delta\Phi$ of the Newton potential. They remain independent at
different times but become correlated at different locations. One can inspect that
(\ref{uncgVT}) is satisfied if we choose the following correlation:
\begin{equation}
\label{corr}
\left\langle\delta\Phi(x,t)\delta\Phi(x',t')\right\rangle 
=\mbox{const.}\times\frac{\hbar G}{|x-x'|}\delta(t-t')~.
\end{equation}  

Thus we have estimated the due uncertainty of the Newton potential (i.e.:
of the space-time geometry in the Newtonian limit). It means an uncertainty
that is present even in empty space. It yields the instability and the
decay of the massive superposition (\ref{superposition}) because it dephases
the two components \cite{diosi1987}. 
The time evolution of $|x_1\rangle$ contains a phase factor
\begin{equation}
\exp\left(-\frac{i}{\hbar}\int_0^t \delta\Phi(x,t')\rho_1(x) d^3x dt'\right)
\equiv e^{-i\chi_1(t)}
\end{equation}
and $|x_2\rangle$ contains $e^{-i\chi_2}$ with $\rho_2$ in place of $\rho_1$.
One forms the expectation value of the squared difference of 
the two phases. Using the correlation (\ref{corr}) yields 
\begin{equation}
\langle(\chi_1(t)-\chi_2(t))^2\rangle=\mbox{const.}\times\frac{E_\Delta}{\hbar}t~,
\end{equation}
where $E_\Delta$ happens to be the expression (\ref{EG}). Therefore the decay 
(i.e.: dephasing) rate of the superposition (\ref{superposition}) coincides 
with (\ref{rate}).
  
\section{Discussion}
\label{Discussion}
Starting point for both of us, as shown in the previous two sections,
was the the inapplicability of standard quantization in general  relativity.
But each of us could implement his concept in the Newtonian limit only.
While Penrose kept, correctly, interpreting the non-relativistic 
proposal in the context of general relativity, I was happy to recognize
that the Newtonian limit is rich and self-contained, though one
should not forget its roots and embedding in general relativity.   

Our  original derivations, outlined in Secs.\ref{Penrose} and \ref{Diosi}, are 
bearing conceptual and even technical similarities as well as important disparities.
The ``uncertainties'' responsible for the decay of massive
superpositions was thought coming  from the ambiguous
Killing vectors  (Sec.\ref{Penrose}) or from  the limited testability 
of the geometry (Sec.\ref{Diosi}). Are the two concepts compatible, 
complementary, or hopelessly contradictory? The answer needs further studies
beyond the scope of the present  work.  
I deliberate on two related things. 

\subsection{Exact derivation}
\label{}
The proposed collapse rate (\ref{rate}) is based on dimensional considerations
hence it contains a numeric constant which is left undefined.
Interestingly, a semiclassical concept and its mathematical realization  
\cite{tilloydiosi2017} by Tilloy and myself confirms the heuristic proposal 
and makes the constant unique. 

The concept looks radically different from that in Sec.\ref{Diosi} but it is 
related to it intrinsically. Assume that the distribution of quantized masses is 
measured everywhere continuously, by hypothetic detectors which are hidden from,
i.e., not part of the physical word. They  are yielding the classical mass 
distribution $\rho$ as the outcomes. This $\rho$ is random, like measurement 
outcomes in quantum systems used to be. The  postulated presence of such universal
and spontaneous measurements serves the coupling of quantized matter to gravity.  
The classical valued $\rho$, used in $\nabla^2\Phi=-4\pi G\rho$, yields the 
classical Newton potential $\Phi$ which is fed back to the Schr\"odinger equation 
of the quantized masses.Now, both the continuous measurement and
the stochastic potential $\Phi$ cause decoherence in the quantized material 
system. Weak (imprecise) measurement causes low decoherence at the price
of high stochasticity of the outcome $\rho$ yielding high decoherence  
by the feedback. At optimum measurement precision the total decoherence 
is the lowest, irrelevant for atomic systems but relevant for massive ones! 
Under it, the superposition (\ref{superposition}) 
will decay exactly at the rate (\ref{rate}) where $E_\Delta$ is defined by 
(\ref{EG}) with the unique prefactor $\mbox{const.}=1/2$. 

\subsection{Footprint of Planck scale uncertainty?}
It would be reassuring to see that the proposed non-relativistic ``uncertainty'',
whether the Killing vector's (Sec.\ref{Penrose}) or the geometry's (Sec.\ref{Diosi}), 
is the non-relativistic limit of the corresponding Planck scale uncertainty.
Penrose \cite{penrose2014} talks about 
\emph{``decay after Planck-scale difference geometry measure''} 
and even conjectures that the decay, according to the formula (\ref{rate}), happens when 
\emph{``two space-times in superposition differ from
one another by an amount of order unity [...] measured in Planck units''}.  
To estimate space-time differences, symplectic measure in linearized gravity is  
mentioned cursorily. However, the whole suggestion about connections to Planck scale is 
missing any quantitative support, be it heuristic or approximate. 

Interestingly, a certain heuristic support existed even before Refs. 
\cite{diosi1986thesis,diosi1987} and was already noticed in them.    
Unruh \cite{unruh1984} proposed unusual (non-canonical) commutators 
between components of the metric tensor $\mathrm{g}$ 
and the Einstein tensor $\mathrm{G}$:
\begin{equation}
\label{commut}
[\mathrm{g}^{\nu\mu}(x),\mathrm{G}^{\rho\sigma}(x')]
=\mbox{const.}\times \ell_{Pl}^2\delta_\nu^\rho \delta_\nu^\sigma \delta^{(4)}(x,x')~,
\end{equation}
with the Planck length $\ell_{Pl}=\sqrt{\hbar G/c^3}$ (here $x,x'$ stand for 
space-time coordinates).
Unruh's motivation was a 
heuristic non-canonical uncertainty relation between the $00$ components 
averaged over four-volume $V^{(4)}$:
\begin{equation}
\delta \mathrm{\bar g}_{00}(x)\delta \mathrm{\bar G}^{00}(x') 
\geq \frac{\ell_{Pl}^2}{V^{(4)}}~,
\end{equation}
now a consequence of (\ref{commut}). The Newtonian limit of this relativistic 
bound leads to the limit (\ref{uncgVT}) of Sec.\ref{Diosi}. We insert 
$\delta\mathrm{\bar g}_{00}=2c^{-2}\delta\bar\Phi$ and $\delta\bar\mathrm{G}^{00}=(1/2)c^{-2}\delta\nabla^2\bar\Phi$, 
as well as $V^{(4)}=cVT.$ Quite remarkably, the velocity of light $c$ cancels:
\begin{equation}
\delta{\bar\Phi}\delta\nabla^2{\bar\Phi}\geq \frac{\hbar G}{VT}~.
\end{equation}
With a deliberate (though justifiable) symmetrization 
$\delta{\bar\Phi}\delta\nabla^2{\bar\Phi}\Rightarrow 
 \delta{\nabla\bar\Phi}\delta\nabla{\bar\Phi}=(\delta{\bar g})^2$,
the obtained uncertainty of the acceleration field 
coincides with (\ref{uncgVT}). 

Much later, and only recently, Ref. \cite{diosi2019} proposed a special 
relativistic construction of conform metric uncertainty around the 
Minkowski metric, whose Newtonian limit confirms the uncertainty (\ref{corr}) 
in Sec.\ref{Diosi}. Introducing a perturbative conformal factor $1 + h$ 
(at $\vert h\vert\ll1$) where the uncertainties $h$ are proportional to 
$\ell_{Pl}$, one  chooses the following relativistic invariant correlation:
\begin{equation}
\langle h(x)h(y) \rangle = \mbox{const.}\times\frac{\ell_{Pl}^2}{(2\pi)^4}
\int e^{-ik(x-y)}\frac{\theta(-k^2)}{-k^2}d^4k~,
\end{equation}
ignoring the issues of regularizing $\theta(-k^2)/k^2$.
One writes $h$ into the form $h = 2\Phi/c^2$, anticipating that $\Phi$ plays 
the role of the (uncertain) Newton potential for non-relativistic matter. 
The limit $c\rightarrow\infty$ does converge, and does exactly to
the expression  (\ref{corr}). 

\section{Closing remarks}
\label{Closing}
For the wider community, interested but not specialized in foundations of physics, 
the conjectured spontaneous collapse rate $E_\Delta/\hbar$ is an easy and only
reference. I focussed on the heuristic derivations of this aesthetic old result.
Hypothesis of spontaneous wave function collapse is a detailed paradigm 
in foundations, with various models, typically unrelated to gravity and
containing fenomenological  parameters\cite{bassi2013}. We should add
that the apparent parameter-independence of our gravity related rate
may be illusory since the rate diverges for pointlike mass distributions.
A fenomenological short-length regulator might be needed. 
Penrose, though, avoids this problem differently\cite{penrose2014}.   
 
``Ah, you are working with Penrose, aren't you?'' --- I was asked a few times.
No, we used to work on our own. There used to be definite parallelisms
and divergences between our struggles in the field of unknowns.     
Roger's interest was a gift. After  many well-defined theoretical tasks that
his talent famously solved,  one earned  his Nobel Prize 2021, 
he challenged the not-so-well-defined problem. 
How is Schr\"odinger's cat bending the space-time?
We  agree that the cat should collapse after a time $\sim\hbar/E_\Delta$. 
We disagree on how this happens, smoothly, suddenly, or some other way. 
The hard task is: gravity-related collapse dynamics that 
conserves energy and momentum. Worth of research. 
If we are on the right track at all ...

\emph{Acknowledgments.} I thank Dr. Maaneli Derakhshani for useful discussions. 
I acknowledge support from the Foundational Questions
Institute and Fetzer Franklin Fund, a donor advised fund of Silicon Valley
Community Foundation (Grant No. FQXi-RFP-CPW-2008),
the National Research, Development and Innovation Office for
"Frontline" Research Excellence Program (Grant No. KKP133827) and research
grant (Grant. No. K12435), the John Templeton Foundation (Grant 62099).

\section{Author declarations}
\subsection{Conflict of Interest}
The authors have no conflicts to disclose.
\subsection{Data Availability Statement} 
The data that support the findings of this study are available from the corresponding 
author upon reasonable request


%
%

%


\bibliography{diosi2021}
\end{document}